# Deep Learning based Segmentation of Optical Coherence Tomographic Images of Human Saphenous Varicose Vein

**Maryam Viqar[1], Violeta Madjarova[1], Amit Kumar Yadav[2], Desislava Pashkuleva[3], Alexander S. Machikhin[4]**

[1]Institute of Optical Materials and Technologies, Bulgarian Academy of Sciences, 109, Acad. G. Bonchev Str., 1113 Sofia, Bulgaria
[2]Department of Electronics Engineering, Aligarh Muslim University, Aligarh, India
[3]Institute of Mechanics, Bulgarian Academy of Sciences, Acad. G. Bonchev St., Block 4, 1113 Sofia, Bulgaria
[4]Scientific and Technological Center of Unique Instrumentation, Russian Academy of Sciences, 15 Butlerova, 117342 Moscow, Russia
Author e-mail address: maryamviqar92@gmail.com

**Abstract:** Deep-learning based segmentation model is proposed for Optical Coherence Tomography images of human varicose vein based on the U-Net model employing atrous convolution with residual blocks, which gives an accuracy of 0.9932. © 2022 The Author(s)

## 1. Introduction

The venous system is important for normal functioning as it accounts for carrying nearly 70% of blood in the human body. The venous diseases like venous insufficiency, varicose veins, vein thrombosis, etc. cause stress and pressure on walls, affects the blood flow, and may even alter the composition of walls. Hence, it's crucial to study the physical properties of the veins for better disease diagnosis and treatment processes [1]. Optical Coherence Tomography (OCT) can generate high resolution images with penetration depth of few millimeters. It can generate 3D structure from stacking of the B-scans. Along with the advancements in the imaging technology, there exists a significant need to develop automated methods to assist the bio-medical processes. Segmentation of veins is important to analyze physical characteristics of the venous walls. It requires high accuracy, as errors may accumulate and lead to incorrect interpretations for the physical properties. In this work, dataset is acquired for the human saphenous vein extracted from a patient suffering from varicose disease. The volume data consists of 800 scans taken on an OCT system. Further, a deep-learning architecture is proposed for segmentation of veins with very high segmentation accuracy. The architecture is U-Net [2] based having residual blocks along with the atrous convolution to generate segmented images separating the inner and outermost boundaries with very high accuracy.

## 2. Saphenous Vein Dataset:

The dataset was acquired using MHz Fourier Domain Mode Lock (FDML) laser OCT system (Optores, Munich, Germany). The system operates at 1.6 MHz sweeping frequency, 1309 nm central wavelength and 100 nm sweeping range providing depth resolution of 15 μm. The lateral resolution of the OCT system was 33 μm and the scanning area was 7 mm x 7 mm (1024 by 1024 points). The saphenous vein of male patient, age 57, suffering from varicose veins was surgically removed at Medical Center "Pirotska", Sofia, Bulgaria. Informed consent was obtained from this patient, prior to surgery. The veins were processed and stored in accordance with the requirements for work with biological samples. The sample veins were prepared at the Institute of Mechanics – Bulgarian Academy of Sciences and experiments were conducted within 24 hours of surgery.

## 3. Segmentation Method:

To segment the saphenous varicose vein, a model based on the U-Net [2] architecture containing Residual blocks [3] along with atrous convolution and dice-loss function is proposed. Symmetrical structure of U-Net has a contraction path responsible for the global features. The expansion path extracts the localized information using convolution and up sampling layers. The fusion connections help to get information of shallow and deep layers for high resolution segmented image. The residual block unit allows easier training of deep networks and can be expressed as:

$$y_j = F(x_j; W_l) + h(x_j); \ x_{j+1} = f(y_j) \qquad (1)$$

Here, for a j-th block, $x_j$ and $x_{j+1}$ represents the input and output respectively. The residual function is $F$, the identity mappings are represented by $h$ and activation function by $f$. The identity mapping properties of residual blocks helps to solve the problem of low convergence rates in deeper networks [3]. Atrous convolution (rate of 2) is used in the bridge block to improve the receptive field for enhanced feature extraction without compromising the resolution. It helps to extract global context information with the same amount of model parameters as in regular convolution. The complete segmentation model is shown in Fig.1a based on U-Net combining residual units [4]. The feature map is refined at the bridge block using the atrous convolution (orange color) for an enhanced receptive field.

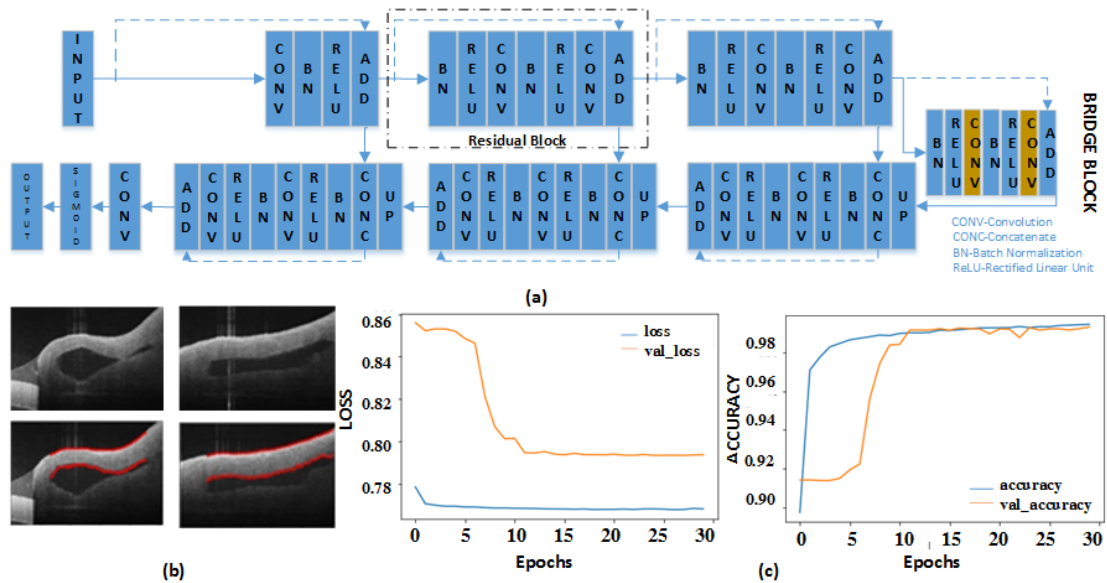

Fig. 1. (a) Segmentation Model (b) Input scan and Segmentation results (c) Plots for accuracy and loss

The encoding and decoding paths both contain three residuals blocks respectively. To obtain the segmented outcome from the feature map, the output of the 7-th block is subjected to 1×1 convolution and Sigmoid activation function. Finally, the dice coefficient loss is used as the loss function. The reason for using the dice coefficient is to solve sample imbalance problem.

Table 1. Results of segmentation on different DL models

|  | ACC | TPR | TNR | Parameters |
|---|---|---|---|---|
| U-Net | 0.9642 | 0.8761 | 0.9912 | 31,031,685 |
| ResUNet | 0.9920 | 0.8656 | 0.9989 | 10,774,213 |
| Proposed | 0.9932 | 0.9321 | 0.9990 | 10,774,213 |

## 4. Experiments, evaluations and conclusions:

The method was evaluated on the Saphenous Vein Dataset introduced in Section 1. It consists of 800 scans with resized resolution of 512×256 pixels. The network was trained on a NVIDIA Quadro P1000 GPU with 4GB graphic memory. The network is trained and tested in Tensorflow 2.0 framework, learning rate of 0.0001 with Adam's optimizer. The model was trained on 600 images with Mini Batch Stochastic Gradient descent, validated and tested on 200 remaining images. Each batch consisted of 16 images. In Table 1 results are compared with the U-Net [2], ResUNet [4] using standard metrics: Acc (Accuracy), True Positive rate (TPR), True Negative rate (TNR). In conclusion, this model achieves the highest accuracy 0.9932, TPR 0.9321 and TNR 0.999 with addition of atrous convolution and dice loss. For visualization, the input (top row) and output (second row) image of the segmentation model is shown in Fig.1b by overlapping the segmented output boundaries on the original image represented using the red line. The plots in Fig.1c shows the corresponding accuracy and loss values for different epochs.

**Acknowledgement:** V.M. thanks National Science Fund of Bulgaria (contract КП-06-Russia/7) and the Russian Foundation for Basic Research (contract 20-58-18007), and European Regional Development Fund within the Operational Programme "Science and Education for Smart Growth 2014–2020" under the Project CoE "National center of Mechatronics and Clean Technologies" BG05M2OP001-1.001-0008. M.V. would like to thank European Union's Horizon 2020 research and innovation programme under the Marie Skłodowska-Curie grant agreement No 956770 for the funding.